\documentclass[twocolumn,showpacs,preprintnumbers,amsmath,amssymb,showkeys]{revtex4}

\usepackage{graphicx}
\usepackage{dcolumn}
\usepackage{bm}

\begin{document}

\title{Complex Acquisition of the Fourier Transform Imaging of an Arbitrary Object}

\author{Minghui Zhang}
\email{zmh@ahu.edu.cn}
\affiliation{Key Laboratory of
Opto-electronic Information Acquisition and manipulation,
 China Ministry of Education,\\ School of Physics and Material Science, Anhui University,
 \\No.3 Feixi Road, Hefei, 230039, P. R. China}

\author{Jianfei Xu}
\affiliation{Key Laboratory of
Opto-electronic Information Acquisition and manipulation,
 China Ministry of Education,\\ School of Physics and Material Science, Anhui University,
 \\No.3 Feixi Road, Hefei, 230039, P. R. China}

\author{Xianfu Wang}
\affiliation{Key Laboratory of
Opto-electronic Information Acquisition and manipulation,
 China Ministry of Education,\\ School of Physics and Material Science, Anhui University,
 \\No.3 Feixi Road, Hefei, 230039, P. R. China}

\date{\today}

\begin{abstract}
A scheme to a complex-valued acquisition of the Fourier transform
imaging was proposed. The main idea is to project the real and the
imaginary parts of a diffraction field to intensity distributions
respectively. The whole procedure was algorithm independent and
needs no \emph{a priori} knowledge of an arbitrary objet. An example
was demonstrated with a numerical modeling and its results.
\end{abstract}

\pacs{42.30.Kq, 42.50.-p, and 42.30.-d}

\keywords{Complex Acquisition; Arbitrary Object; Fourier transform
imaging.} \maketitle

When talk about the image forming, the process by instruments like
the eye, the camera, the reflecting and refracting telescope, and
the microscope etc., were generally refers to a point to point
correspondence between two real spaces \cite{1}. If a situation
requires full information about the objects rather than only a
magnitude transmittance or reflectance, such procedures would be
invalid because what they mapped was only intensity relations, and
phase information would thus lost. A possible way to obtain the
phase knowledge of the object's transmittance is to convert it into
a spatially varying diffraction pattern with form of Fourier
transform (supposed in Fraunhofer region), and then, invert it to
its object function. This procedure is based on the fact that
Fourier transform keeps unitary relations between real and
reciprocal spaces. Unfortunately, the diffraction fields are also
only recordable by an intensity-sensitive detector and this awkward
fact would lead phase loss to occur in the reciprocal space again.
To solve the phase problem, efforts have been paid by utilizing
oversampling methods \cite{2} with iterative algorithms
\cite{3,4,5}. The modulus of Fourier transform can thus be phased
and then inverting it into an object functions. This method was
reported recently in X-ray diffraction microscopy \cite{6} and has
been extended from x-ray crystallography \cite{7}to the imaging of
noncrystalline materials \cite{8}, and single cells \cite{9}. Its
potential for imaging of single protein complexes by using
ultra-short X-ray pulses with extreme intensity were also discussed
\cite{10}. The excellent works as mentioned above have now achieved
the imaging of single virion even with a resolution of $22 nm$
\cite{11}.

Anyway, the phase-less magnitude alone would not be able to sustain
the unitary property of Fourier transform. Therefore, iterative
algorithms to recover the phase encoded in the diffraction pattern
have to rely on \emph{a priori} knowledge of the objects more or
less. If a object was arbitrary \emph{i.e. }not purely absorptive or
not purely phased, the ambiguity would arise \cite{12}. The
difficulties of the complex-valued acquisitions without any original
information seemed to have already been predestined by quantum
mechanics. \emph{i.e}., the complex amplitude can not be specified
exactly in a single measurement \cite{13}. On the other hand, for
coherent sources are not obtainable in wave length like hard $X$-ray
region, potential advantages \cite{14} might be jeopardized when
using the ultra-short, intense $X$-ray pulses\cite{10}.

In this paper, we theoretically propose a methods and a scheme to a
complex-valued acquisition of the Fourier transform imaging of the
arbitrary object's transmittance with incoherent light. The main
thought is to project the real and the imaginary parts of a
Fourier-transformed field to intensity distributions respectively.
The whole procedure is algorithms free and needs no \emph{a priori}
knowledge of an arbitrary objet.
\begin{figure}
\centerline{\includegraphics* [bb=102 135 293 206,scale=1.2]
{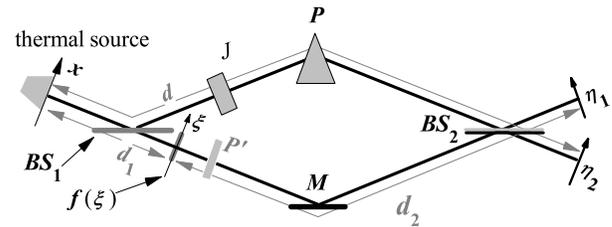}}\caption{\label{fig:set_up}The proposed scheme for set
up }
\end{figure}

The proposed scheme for set up is shown in Fig.\ref{fig:set_up}.
Fields from the thermal source is split by a $50/50$ beam splitter
$BS_{1}$ to form a two-arm optical system with equal distance from
source plane $x$ to planes $\eta_{1}$, and $\eta_{2}$, where
intensity information will be registered. At the cross section of
both arms before $\eta_{1}$, and $\eta_{2}$, the other $50/50$ beam
splitter $BS_{2}$ would insert. As for $BS_{2}$, we must manage to
ensure that the half wave loss only occurs when the light reflects
only on one of its surfaces. When a situation needs to introduce a
phase factor shift of $j$, a phase plate $J$ would be inserted into
the upper part of the system. The total scheme differs mainly from
the Mach-Zehnder interferometer is that it uses a prismatic lens $P$
rather than a plane mirror to guide the optical path in its upper
part. The object with complex transmittance of $f(\xi)$ is placed at
plane $\xi$ in lower part of the setup. The distance from plane
$\xi$ to plane $x$, and to plane $\eta_{1}$, and from plane $x$ to
plane $\eta_{2}$ are $d_{1}$, $d_{2}$ and $d$ respectively. The
equal length of the two arms requires that
\begin{align}
d = d_1  + d_2.\label{eq:equal_length}
\end{align}
Among the setup, an optional phase plate $P'$ might insert at the
lower part of the scheme to form a fixed optical path difference
between two arms of a phase factor of $1/\sqrt{j}$, whereas, it was
not a necessary component.

Although chaotic light fields fluctuates randomly, the relations
among instantaneous values of optical fields $E$ in plane
$\eta_{1}$, $\eta_{2}$ and $\xi$ are deterministic according to
Fresnel diffraction theory. To illustrate the theoretical bases for
this complex-valued retrieval procedure clearly, first we suppose,
that we did not facilitate the beam splitter $BS_{2}$, phase plate
$P'$, and $J$ in Fig.\ref{fig:set_up}. In this situation, the
Optical fields on $\eta_{1}$, and $\eta_{1}$ fulfill \cite{15}
\begin{align}
E\left( {\eta _2 } \right) = \frac{{e^{jkd} }} {{j\lambda
d}}\int\limits_x {E\left( x \right)e^{jk\frac{{\left( {\eta _2  - x}
\right)^2 }} {{2d}}} dx},\label{eq:fresnel_int_1}
\end{align}
and
\begin{align}
E\left( {\eta _1 } \right) = \frac{{e^{jkd_2 } }} {{j\lambda d_2
}}\int\limits_\xi  {\left[ {\frac{{e^{jkd_1 } }} {{j\lambda d_1
}}\int\limits_x {E\left( {x'} \right)e^{jk\frac{{\left( {\xi  - x'}
\right)^2 }}
{{2d_1 }}} dx'} } \right]}  \nonumber \\
  \begin{array}{*{20}c}
\end{array}  \times f\left( \xi  \right)e^{jk\frac{{\left( {\eta _1  - \xi } \right)^2 }}
{{2d_2 }}} d\xi \label{eq:fresnel_int_2}
\end{align}
under the Fresnel approximation. Let the thermal source is totally
chaotic \cite{16} with evenly distributed intensity, \emph{i.e.},
\begin{align}
\left\langle {E\left( x \right)E\left( {x'} \right)} \right\rangle =
I\delta \left( {x - x'} \right),\label{eq:chaotic}
\end{align}
in which  $\langle\cdot\cdot\cdot\rangle$  stands for the assemble
average. Then, the derivation of the mutual intensity between plane
$\eta_{1}$ and $\eta_{2}$ can be derived from Eq.
(\ref{eq:equal_length}) to Eq.(\ref{eq:chaotic}) to be
\begin{align}
  \left\langle {E\left( {\eta _1 } \right)E^* \left( {\eta _2 } \right)} \right\rangle  = \sqrt j e^{j\frac{\pi }
{\lambda }\left( {\frac{{\eta _1^2  - \eta _2^2 }} {{d_2 }}}
\right)} \frac{I}
{{\pi \sqrt {\lambda ^5 dd_1 d_2^3 } }} \nonumber \\
   \times F\left( {\frac{{\eta _1  - \eta _2 }}
{{\lambda d_2 }}} \right) \propto \sqrt j e^{j\frac{\pi } {\lambda
}\left( {\frac{{\eta _1^2  - \eta _2^2 }} {{d_2 }}} \right)}
{\mathcal{F}}\left\{ {f\left( \xi  \right)} \right\}.
\label{eq:fourier}
\end{align}
In which $F\mathcal{F}(\cdot\cdot\cdot)$ refers to
$\mathcal{F}\{f(x)\}$,the Fourier transform of $f(x)$. Now with the
beam splitter $BS_{2}$ facilitated, the fields in plane $\eta_{1}$,
and $\eta_{2}$ turn out to be:
\begin{align}
E_1 \left( {\eta _1 } \right) = \frac{1} {{\sqrt 2 }}\left[ {E\left(
{\eta _1 } \right) - E\left( {\eta _2 } \right)}
\right],\label{with_bs2_1}
\end{align}
and
\begin{align}
E_2 \left( {\eta _2 } \right) = \frac{1} {{\sqrt 2 }}\left[ {E\left(
{\eta _1 } \right) + E\left( {\eta _2 } \right)}
\right],\label{with_bs2_2}
\end{align}
when half wave loss only on one side of $BS_{2}$ was considered.
Their corresponding intensity distribution registered on plane
$\eta_{1}$, and $\eta_{2}$ are
\begin{align}
  I_1 \left( {\eta _1 } \right) = \left\langle {E_1 \left( {\eta _1 } \right)E_1^* \left( {\eta _1 } \right)} \right\rangle  = \frac{1}
{2}\left( {\left| {E\left( {\eta _1 } \right)} \right|^2 } \right. \nonumber\\
  \left. { - 2\operatorname{Re} \left\langle {E\left( {\eta _1 } \right)E^* \left( {\eta _2 } \right)} \right\rangle  + \left| {E\left( {\eta _2 } \right)} \right|^2 } \right)
,\label{intensity_1}
\end{align}
and
\begin{align}
  I_2 \left( {\eta _2 } \right) = \left\langle {E_2 \left( {\eta _2 } \right)E_2^* \left( {\eta _2 } \right)} \right\rangle  = \frac{1}
{2}\left( {\left| {E\left( {\eta _1 } \right)} \right|^2 } \right. \nonumber \\
  \left. { + 2\operatorname{Re} \left\langle {E\left( {\eta _1 } \right)E^* \left( {\eta _2 } \right)} \right\rangle  + \left| {E\left( {\eta _2 } \right)} \right|^2 } \right)
,\label{intensity_2}
\end{align}
respectively. The fact can be easily seen from two equations above
that the real part of Eq.(\ref{eq:fourier}) is embedded in the
intensity registration on both plane $\eta_{1}$, and $\eta_{2}$, and
can be extracted by subtracting the two equal backgrounds by means
of
\begin{align}
\operatorname{Re} \left\langle {E\left( {\eta _1 } \right)E^* \left(
{\eta _2 } \right)} \right\rangle  = \frac{{I_2 \left( {\eta _2 }
\right) - I_1 \left( {\eta _1 } \right)}} {2}.\label{real}
\end{align}
Further more, the linearity Eq.(\ref{eq:fourier}) indicates that if
a phase shift of $\phi$ is introduced in the upper or the lower part
of the scheme shown as Fig.\ref{fig:set_up}, a phase factor of
$e^{-j\phi}$ or $e^{j\phi}$ would be multiplied on the right side of
the equation consequently. Following the thought, if we inserted a
phase plate $J$ into the upper part of the system to introduce a
phase shift of $\pi/2$, the complexes vector stands for
Eq.(\ref{eq:fourier}) would rotate an angle of $-\pi/2$
consequently. So the imaginary part of Eq.(\ref{eq:fourier}) can
also be retrieved in a way similar to eq.(\ref{real}) by:
\begin{align}
\operatorname{Im} \left\langle {E\left( {\eta _1 } \right)E^* \left(
{\eta _2 } \right)} \right\rangle  = \frac{{I'_2 \left( {\eta _2 }
\right) - I'_1 \left( {\eta _1 } \right)}} {2},\label{eq:imaginary}
\end{align}
since a factor of $-j$ had been brought into Eq.(\ref{eq:fourier}).
In Eq.(\ref{eq:imaginary}), $I'_{2}(\eta_{2})$ and
$I'_{1}(\eta_{1})$ are intensity registration on plane $\eta_{2}$,
and $\eta_{1}$, after a phase plate $J$ was inserted into the upper
part of the system as Fig.\ref{fig:set_up} shows. Note that we use a
prismatic lens $P$ rather than a plane mirror to guide the optical
path in upper part of the system. This arrangement leads to a
bilateral symmetry between coordinates in plane $\eta_{1}$, and
$\eta_{2}$, \emph{i.e.},
\begin{align}
\eta _1  =  - \eta _2 \left( { = \eta } \right).\label{eq:symmetry}
\end{align}
Comparing Eq.(\ref{eq:fourier}), Eq.(\ref{real}),
Eq.(\ref{eq:imaginary}), and Eq.(\ref{eq:symmetry}), we proposed the
complex-valued acquisition of an object's Fourier transform imaging.
The procedure can be written in one equation as:
\begin{align}
  F\left\{ {f\left( \xi  \right)} \right\} = F\left( {\frac{{2\eta }}
{{\lambda d_2 }}} \right) \propto \frac{1} {{\sqrt j }}\left(
{\frac{{I_2 \left( {\eta _2 } \right) - I_1 \left( {\eta _1 }
\right)}}
{2}} \right. \nonumber \\
  \left. { + j\frac{{I'_2 \left( {\eta _2 } \right) - I'_1 \left( {\eta _1 } \right)}}
{2}} \right) .\label{eq:all_in_one_eq}
\end{align}

\begin{figure}
\centerline{\includegraphics* [bb=17 81 364 188,scale=0.7]
{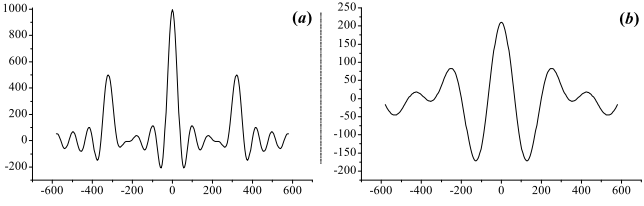}}\caption{\label{fig:fourier}The real (a) and imaginary
(b) parts of the Fourier transform as a function of the space
coordinates $\eta$. It originates from a complex-valued
transmittance (eq.(14)) of a conceived object.}
\end{figure}

To give an example of the retrieval, we conceived an object with a
complex-valued transmittance of:
\begin{align}
  f\left( \xi  \right) = \left\{ {\left( {{\text{1 + cos0}}{\text{.05}}\xi } \right) + j\left[ {rect\left( {\frac{{\xi  + 150}}
{{105}}} \right)} \right.} \right. \nonumber \\
  \left. {\left. { + rect\left( {\frac{{\xi  - 150}}
{{105}}} \right)} \right]} \right\}rect\frac{\xi } {{1000}}
.\label{conceived_obj}
\end{align}
In which 0.05, 150, 105, and 1000 are space parameters with $\mu m$
unit in plane $\xi$. If the object were to be illuminated by a
coherent light with a wavelength of $\lambda = 0.532\mu m$, the real
and imaginary parts of its Fourier transform as a function of space
coordinates $\eta$ would be:
\begin{align}\operatorname{Re}\left( {{\mathcal{F}}\left\{ {f\left( \xi  \right)} \right\}} \right) = 500\sin c\left[ {1000\left( {\frac{\eta }
{{\lambda d_2 }} + \frac{{0.05}}
{{2\pi }}} \right)} \right] \nonumber \\
   + 1000\sin c\left( {1000\frac{\eta }
{{\lambda d_2 }}} \right) \nonumber \\
   + 500\sin c\left[ {1000\left( {\frac{\eta }
{{\lambda d_2 }} - \frac{{0.05}} {{2\pi }}} \right)} \right]
,\label{real_numerical}
\end{align}
and
\begin{align}
\operatorname{Im} \left( {{\mathcal{F}}\left\{ {f\left( \xi \right)}
\right\}} \right) = 210\sin c\left( {105\frac{\eta } {{\lambda d_2
}}} \right)\nonumber\\\times\cos \left( {300\pi \frac{\eta }
{{\lambda d_2 }}} \right);\label{imaginary_numerical}
\end{align}
as Fig.\ref{fig:fourier}shows. In Eq.(\ref{real_numerical}) and
Eq.(\ref{imaginary_numerical}), $d_{2} =75,000 \mu m$, refers to a
distance the coherent light field with $\lambda=0.532 \mu m$ from
the object propagates until it reaches the plane $\eta$.

Based on the theory of statistical optics, we numerically modeled
the dynamic process of the whole retrieval procedure under the setup
scheme of Fig.\ref{fig:set_up}, by using the conceived object with
transmittance of Eq. (\ref{conceived_obj}). An optional phase plate
prism $P'$ was inserted at the lower part of the scheme to form a
fixed optical path difference of $\pi/4$ between two arms to
compensate the factor of $1/\sqrt{j}$ in Eq.(\ref{eq:fourier}).

In the modeling, the monochromatic thermal light featured circular
Gaussian random process with zero mean \cite{16}, the wave length of
the thermal light was selected to be $\lambda = 0.532 \mu m$, the
propagation of the fields among the setup fulfills the Fresnel
diffraction integrations stated by Eq.(\ref{eq:fresnel_int_1})to
Eq.(\ref{eq:fresnel_int_2}). For the setup, $d_{1}$, $d_{2}$ and $d$
are set to $60,000 \mu m$, $75,000 \mu m$, and $135,000 \mu m$.
After an accumulative intensity registration, which covers $20,000$
times of independent coherent time, the averaged intensities in both
plane $\eta_{1}$ and $\eta_{2}$ were brought about in FIG. 3. In
which (a) and (a') are intensity registered in plane $\eta_{2}$
before and after a phase plate of $\pi/2$, \emph{i.e.},$J$ was
inserted; (b) and (b') are intensity registered in plane $\eta_{1}$
before and after a phase plate of $J$ was inserted; (c) and (c') was
the intensity difference between plane $\eta_{2}$ and $\eta_{1}$
before and after $J$ was inserted. Comparing (c) and (c') with (a)
and (b) in Fig.\ref{fig:fourier}, the numerical results shows that
the real and imaginary parts of the Fourier imaging of a conceived
object (Eq.(\ref{conceived_obj})) were both retrieved in a double
scale coordinate of Fig.\ref{fig:fourier}.
\begin{figure}
\centerline{\includegraphics* [bb=75 14 427 300,scale=0.7]
{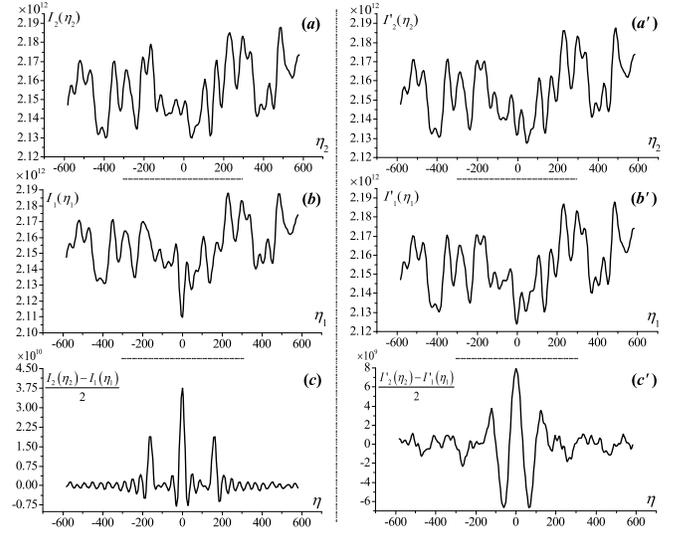}}\caption{\label{fig:results}Numerical results for the
Complex-valued Retrieval of a conceived object. Its left and right
column concerns intensity registration before and after a phase
plate $J$ of $\pi/2$ was inserted in the upper part of scheme in
Fig.\ref{fig:set_up}. Comparing (c) and (c') with (a) and (b) in
FIG. \ref{fig:fourier}, the numerical results shows that the real
and imaginary parts of the Fourier imaging of a conceived object
(Eq.(\ref{conceived_obj})) were both retrieved in a sub-wave length
scale.   }
\end{figure}

In summary, by using incoherent light, we theoretically proposed a
scheme to a complex acquisition of the Fourier transform imaging.
The whole procedure needs no \emph{a priori} knowledge of an
arbitrary object. The unitary property of the Fourier transform thus
sustains by intensity recordable detectors. In recent years,
coincidence imaging by classical thermal sources has been widely
investigated both in real \cite{17, 18} and reciprocal spaces
\cite{19,20,21}. In these works, images were carried out either by
two photon coincident rate or by correlation functions of intensity
fluctuations. They both stand for the modules of optical fields.
Useful methods to retrieval a complex-valued objects with discrete
variables \cite{22,23}  were also reported combined with
oversampling methods \cite{2} and iterative algorithms \cite{3,4,5}.
The authors of Ref.\cite{12} reconstruct the complex object without
\emph{a priori} knowledge, but still based on computer operations.
Unlike these works, we map the complex valued object in real space
to complex valued Fourier transform only by intensity registration
\cite{wang_prl_2009}. As the unitary property remains, the retrieval
procedure by this scheme equals to imaging an arbitrary complex
valued object in a full sense. To our knowledge, it is the first
physical proposal to sustain the unitary property of an arbitrary
object when imaging without \emph{a priori} knowledge.

The authors would like to thank 211 Project Foundation of Anhui
University with number of 02203104.

\end{document}